\documentclass[review]{elsarticle}

\pdfoutput=1

\usepackage[utf8]{inputenc}

\usepackage{booktabs}
\usepackage{wrapfig}
\usepackage{graphicx}
\usepackage{pdfpages}

\usepackage{url}
\usepackage{breakurl}
\usepackage[breaklinks]{hyperref}

\usepackage{lineno,hyperref}
\usepackage{amsmath}

\modulolinenumbers[5]

\journal{Journal of Epidemiology, Elsevier}

\begin{document}

\begin{frontmatter}

\title{A model of COVID-19 pandemic evolution in African countries}
%\title{Elsevier \LaTeX\ template\tnoteref{mytitlenote}}

%\author{Ketevi Assamagan\fnref{myfootnote}}
%\address{Brookhaven National Laboratory Upton, New York, USA}
%\fntext[myfootnote]{Since 2004.}
%
%\author{Somiealo  Azote\fnref{myfootnote}}
%\address{Université de Lomé, Physics Department, Lomé, Togo}
%%\fntext[myfootnote]{Since 2020.}
%
%\author{Toivo Mabote\fnref{myfootnote}}
%\address{Medical Physics, Department at the University Eduardo Mondlane, Mozambique}
%%\fntext[myfootnote]{Since 2020.}
%
%\author{Kondwani C. C Mwale\fnref{myfootnote}}
%\address{University of Rwanda-African Center of Excellence for Innovative Teaching and learning Mathematics/ Science, Rwanda}
%%\fntext[myfootnote]{Since 2020.}
%
%\author{ George Zimba\fnref{myfootnote}}
%\address{University of Jyväskylä,Finlande}
%%\fntext[myfootnote]{Since 2020.}
%
%\author{Ebode Onyie\fnref{myfootnote}}
%\address{The University of Yaounde I, Faculty of Science, Department of Physics, Yaounde, Cameroon}
%%\fntext[myfootnote]{Since 2020.}


\author[add1]{Kossi Amouzouvi}
\author[add2]{K\'et\'evi A. Assamagan\corref{cor1}}
\ead{ketevi@bnl.gov}
\author[add3]{Somi\'ealo  Azote}
\author[add4]{Simon H. Connell}
\author[add5]{Jean Baptiste Fankam Fankam}
\author[add6]{Fenosoa Fanomezana}
\author[add7]{ Aluwani Guga}
\author[add8]{Cyrille E. Haliya}
\author[add9]{Toivo S. Mabote}
\author[add9]{Francisco Fenias Macucule}
\author[add10]{Dephney Mathebula\corref{cor1}}
\ead{dephneymathebula@yahoo.com}
\author[add7]{Azwinndini Muronga}
\author[add11]{Kondwani C. C. Mwale}
\author[add12]{Ann Njeri}
\author[add5]{Ebode F. Onyie}
\author[add13]{Laza Rakotondravohitra}
\author[add14]{George Zimba}

\cortext[cor1]{Corresponding Authors}

\address[add1]{Kwame Nkrumah University of Science and Technology, Ghana}
\address[add2]{Brookhaven National Laboratory, Physics Department, Upton, New York, USA}
\address[add3]{Universit\'e de Lom\'e, D\'epartement de Physique, Lom\'e, Togo}
\address[add4]{University of Johannesburg, Johannesburg, South Africa}
\address[add5]{University of Yaounde I,  Department of Physics,Yaounde, Cameroon}
\address[add6]{University of Antananarivo, Madagascar}
\address[add7]{Nelson Mandela University, South Africa}
\address[add8]{University of Abomey-Calavi, International Chair in Mathematical Physics and Applications, Cotonou, Benin}
\address[add9]{Universidade Eduardo Mondlane, Grupo de Astrofísica e Ci\^{e}ncias Espaciais, Maputo, Mozambique}
\address[add10]{University of Venda, Applied Mathematics and Mathematics Department, South Africa}
\address[add11]{University of Rwanda, African Center of Excellence for Innovative Teaching and Learning Science, Kigali, Rwanda}
\address[add12]{University of Manchester, UK}
\address[add13]{Duke University Medical Center, USA}
\address[add14]{University of Jyv\"{a}skyl\"{a}, Department of Physics, Jyv\"{a}skyl\"{a}, Finland}

%\today
\begin{abstract}
We studied the COVID-19 pandemic evolution in selected African countries. For each country considered, we modeled simultaneously the data of the active, recovered and death cases. In this study, we used a year of data since the first cases were reported. We estimated the time-dependent basic reproduction numbers, $R_0$, and the fractions of infected but unaffected populations, to offer insights into containment and vaccine strategies in African countries. We found that $R_0\leq 4$ at the start of the pandemic but has since fallen to $R_0 \sim 1$. The unaffected fractions of the populations studied vary between $1-10$\% of the recovered cases.
\end{abstract}

\begin{keyword}
COVID-19 \sep SIDARTHE \sep Basic Reproduction Number, SARS-CoV-2
%\texttt{elsarticle.cls}\sep \LaTeX\sep Elsevier \sep template
%\MSC[2010] 00-01\sep  99-00
\end{keyword}

\end{frontmatter}

%\linenumbers
%The project is proposed by a collaborative team composed of experienced physicists and
%research scientists with expertise in mathematical modeling, data science, computational simulations,
%and analysis tools. The team also contains several African nationals with the experience to collect data
%from their countries and to understand the conditions under which the data were collected.
%\newpage
\section{Introduction}
\label{sec:intro}
\par\noindent SARS-CoV-2 is the virus responsible for COVID-19, a coronavirus disease that has caused a worldwide pandemic~\cite{worldhth2020}. Infected persons may show symptoms of respiratory illnesses and most people recover without serious medical interventions. However, complications and risk of death may occur in older folks and those with pre-existing medical conditions~\cite{Worldh2021}. SARS-CoV-2 spreads through oral or nasal discharges from infected persons. Worldwide, upwards of a hundred million confirmed cases and over two million deaths have been registered~\cite{Covid19WorldMap}. To reduce transmissions, hygienic measures, social distancing and quarantines have been adopted and travel restrictions imposed, with severe impacts on the world economies~\cite{Worldh2021}. Africa has had over three million and a half cases of COVID-19 with about ninety thousand deaths---with close to fifty percent of the cases in South Africa~\cite{AfricaCDC2021}. 

In the current uncertain situation, we have carried out a mathematical modeling of COVID-19 data from Africa countries to estimate their time-dependent basic reproduction numbers. The current study is based on COVID-19 data of nine African countries, namely Cameroon, Ghana, Kenya, Madagascar, Mozambique, Rwanda, South Africa, Togo, and Zambia. The data used cover periods of twelve months from the first cases of COVID-19 in the countries considered. The study presented here is an extension of an earlier one done on the first few months of the COVID-19 pandemic~\cite{assamagan2020study}.

We used the SIDARTHE model~\cite{Giordano2020} to study the evolution of the COVID-19 pandemic in the aforementioned countries and estimated their time-dependent $R_0$---an epidemiological parameter often used to gauge the evolution of a pandemic. Estimation of $R_0$ considers various biological, socio-economic, environmental and behavioral factors~\cite{Giordano2020}; however, it is model dependent. The $R_0$ may be used to estimate the fraction of a population to vaccinate. A number of vaccines have been developed, offering hope to bring the pandemic under control. However, emergence of new strains---because of mutations of the virus---might reduce effectiveness of the vaccines~\cite{COVAX2021}. 

The paper is organized as follows. In Section~\ref{sec:mod}, we discuss the SIDARTHE model and briefly review the basic reproduction number in Section~\ref{sec:r0}. In Section~\ref{sec:stra}, we present the analysis and results. In Section~\ref{sec:disc}, we discuss the implications of the results, and we offer concluding remarks in Section~\ref{sec:conc}.

\section{SIDARTHE Model Formulation}
\label{sec:mod}
\noindent The SIDARTHE model considers eight stages of pandemic evolution, namely $S$, susceptible (uninfected); $I$, infected (asymptomatic or pauci-symptomatic infected, undetected); $D$, diagnosed (asymptomatic infected, detected); $A$, ailing (symptomatic infected, undetected); $R$, recognized (symptomatic infected, detected); $T$, threatened (infected with life-threatening symptoms, detected); $H$, healed (recovered) and $E$, extinct (dead)~\cite{Giordano2020}. For detailed derivation of the SIDARTHE model, refer to Ref.~\cite{Giordano2020}. Graphical representations of the stages of pandemic evolution in the SIDARTHE model are shown in Refs~\cite{assamagan2020study, Giordano2020}; the model proposes the following system of ordinary differential equations to describe the time evolution of a pandemic~\cite{Giordano2020}.
%\begin{figure}[!htb]
% 	\centering
% 	\includegraphics[width=12cm, height=9cm]{UpdateSIDARTHE_Drawing}
%  	\caption{Flow chart representing the interactions among different stages of pandemic %evolution in the SIDARTHE model~\cite{Giordano2020}.}
%   	\label{fig:Model}
% \end{figure}
 \begin{eqnarray}
 \left\{\begin{array}{lcl}
 \dot{S}(t) ~=~ -S(t)[ \alpha I(t) + \beta D(t) + \gamma A(t) + \delta R(t)], 	\\\\
 \dot{I}(t) ~=~ S(t)[\alpha I(t) + \beta D(t) + \gamma A(t) + \delta R(t)]-(\epsilon +\lambda +\zeta)I(t),   \\\\
 \dot{ D}(t) ~=~ \epsilon I(t)-(\eta +\rho)D(t),\\\\
 \dot{A}(t) ~=~  \zeta I(t)-(\theta +\mu)+\kappa)A(t),\\\\
 \dot{R}(t) ~=~ \eta D(t) + \theta A(t)-(\nu+\chi)R(t),\\\\
 \dot{T}(t) ~=~ \mu A(t)+\nu R(t)-(\sigma +\tau)T(t),\\\\
  \dot{H}(t) ~=~ \lambda I(t) +\rho D(t) + \kappa A(t) + \chi R(t) + \sigma T(t) ,\\
\dot{E}(t) ~=~\tau T(t)
\label{EQN:101} 
 \end{array}\right.
 \end{eqnarray}

In Equation~\ref{EQN:101}, the expression for the healed population, $\dot{H}(t)$, contains the term $\lambda I(t)$: this is the fraction of the population that recovered from the infected category, $I(t)$, without symptoms nor diagnosis---we denote this term, $\lambda I(t)$, the "unaffected population". The modeling of the recovered population contains the unaffected population; however, as discussed later in Sections~\ref{sec:stra} and~\ref{sec:disc}, the collected data do not account for the unaffected categories. This is an important feature of the model. A large value of the "unaffected population" will demonstrate the lack of testing.

\section{Basic Reproduction number}
\label{sec:r0}
\noindent  The basic reproduction number plays an important role in mathematical epidemiology; it is the average number of secondary cases produced by an infected individual in a population where everyone is susceptible~\cite{VdD2002}. In the SIDARTHE model, $R_0$ can be derived as:
\begin{eqnarray}
\label{eq:r0}
R_0 = \frac{\alpha}{r_2}+ \frac{\beta \epsilon}{r_1 r_2}+ \frac{\delta \nu \epsilon}{r_1r_2r_4} + \frac{\delta \zeta \theta }{r_1 r_3 r_4}\end{eqnarray}
where $r_1 = \epsilon + \zeta + \lambda, ~r_2 = \eta + \rho, ~ r_3 = \theta + \mu + \kappa, ~r_4 = \nu + \xi$. Ref.~\cite{Giordano2020} shows details on the $R_0$ derivation. As shown in Equation~\ref{EQN:101}, $R_0$ depends on the model parameters that affect pandemic evolution. The aim of this analysis was to estimate $R_0$ with model parameters that describe the real data. Therefore, it is important to understand the model parameters and to make sure they are extracted correctly.

\section{Analysis of COVID-19 Data}
\label{sec:stra}

\noindent In this section, we discuss the analysis strategy. For each country, we modeled the three datasets of active, recovered and death cases by finding the SIDARTHE model parameters that best match the time-evolution of the data. Then, the basic reproduction number is computed with the best-matched parameters using Equation~\ref{eq:r0}. Depending on the evolution of the pandemic and the response measures applied, the best-matched parameters may change over time, thus providing an evolution of the basic reproduction number as a function of time. We propagated to the $R_0$ estimates of the statistical uncertainties related to the numbers of tests done and cases identified. We also applied a systematic uncertainty to the $R_0$ estimates based on the fraction of the infected but unaffected population; this fraction is predicted by the model; however, it not accounted for in the data. In the following subsections, we discuss the data analysis and results for each country studied.

\subsection{Analysis of COVID-19 data of South Africa}
\noindent The National Institute for Communicable Diseases confirmed the first COVID-19 case on March 5, 2020, which arose due to a returning traveler 
from Italy~\cite{NICD2020}.
%The patient was a 38 years old male from the Kwazulu-Natal region; he had traveled to Italy with his wife. The couple was part of a group of ten individuals who traveled to Italy and returned to South Africa on March 1, 2020~\cite{NICD2020}. 
\noindent By March 13, 2021, South Africa had a total of $1528414$ COVID-19 cases, making it the African country with the highest number of COVID-19  infections~\cite{owidcoronavirus}. 

\noindent Over the course of the first wave, the South African government implemented five different levels of respective lockdown regulations with alert level one being the most relaxed and alert level five the strictest~\cite{greyling2021good}. 
%The levels were implemented to reduce the spread of COVID-19 and give time to South Africa’s health system to prepare itself~\cite{greyling2021good}. 
Table~\ref{table:1} shows a summary of the lockdown levels with their periods of implementation. 
\begin{table}[!h]
\centering
\begin{tabular}{|p{2cm}|p{3cm}|p{7cm}|}
%\begin{tabular}{|l|l|}
\hline
 \textbf{ Level}   &  \textbf{Period} & Summary \\ \hline
Adjusted alert level 3 & Effect from 29 December 2020 to 28 February 2021 & Gatherings and recreational activities severely restricted. Curfew imposed and restrictions on alcohol sales and consumption.  Restricted cross border travels were allowed.\\ \hline
Alert level 1& Effect from 21 September to 28 December 2020  & International travel allowed with restrictions, domestic travel open, all economic sectors open. For this and all lockdown levels, restrictions or closure on sporting events, religious gatherings, public entertainment, restaurants and similar. \\ \hline 
Alert level 2 &  Effect from 18 August to 20 September 2020 & Domestic air and road travel restored. Further economic sectors opened\\ \hline
Alert level 3 & Effect from 1 June to 17 August 2020 & Economy more open than Level 4, but still restrictions, example, no restaurants, restricted inter-provincial travel.\\ \hline
Alert level 4 & Effect from 1 to 31 May 2020 & Some non-essential services operate, with restrictions, eg: agriculture, mining, communications, business travel. Local travel within curfew, restricted provincial travel. \\ \hline
Alert level 5 & Effect from midnight 26 March to 30 April 2020 & Only essential services, transport and movement restrictions\\ \hline
\end{tabular}
\caption{Lockdown levels implemented by the South African government since March 26, 2020 \cite{Levels2020}.}
\label{table:1}
\end{table}
%On March 26, 2020, the country went into the hardest lockdown in which the government put drastic measures to contain the spread of the virus and save lives. Borders were closed; no international travels; no inter-provincial travels; no alcohol and cigarette sales. All businesses closed except those offering essential services such as supermarkets and pharmaceuticals. It was required to work remotely, except essential service workers such as health workers. People were allowed outside to get essentials only when absolutely necessary. There was a night curfew.  This went on until April 30, 2020. 

\begin{figure}[!h]
 \begin{center}
 	\includegraphics[width=\textwidth]{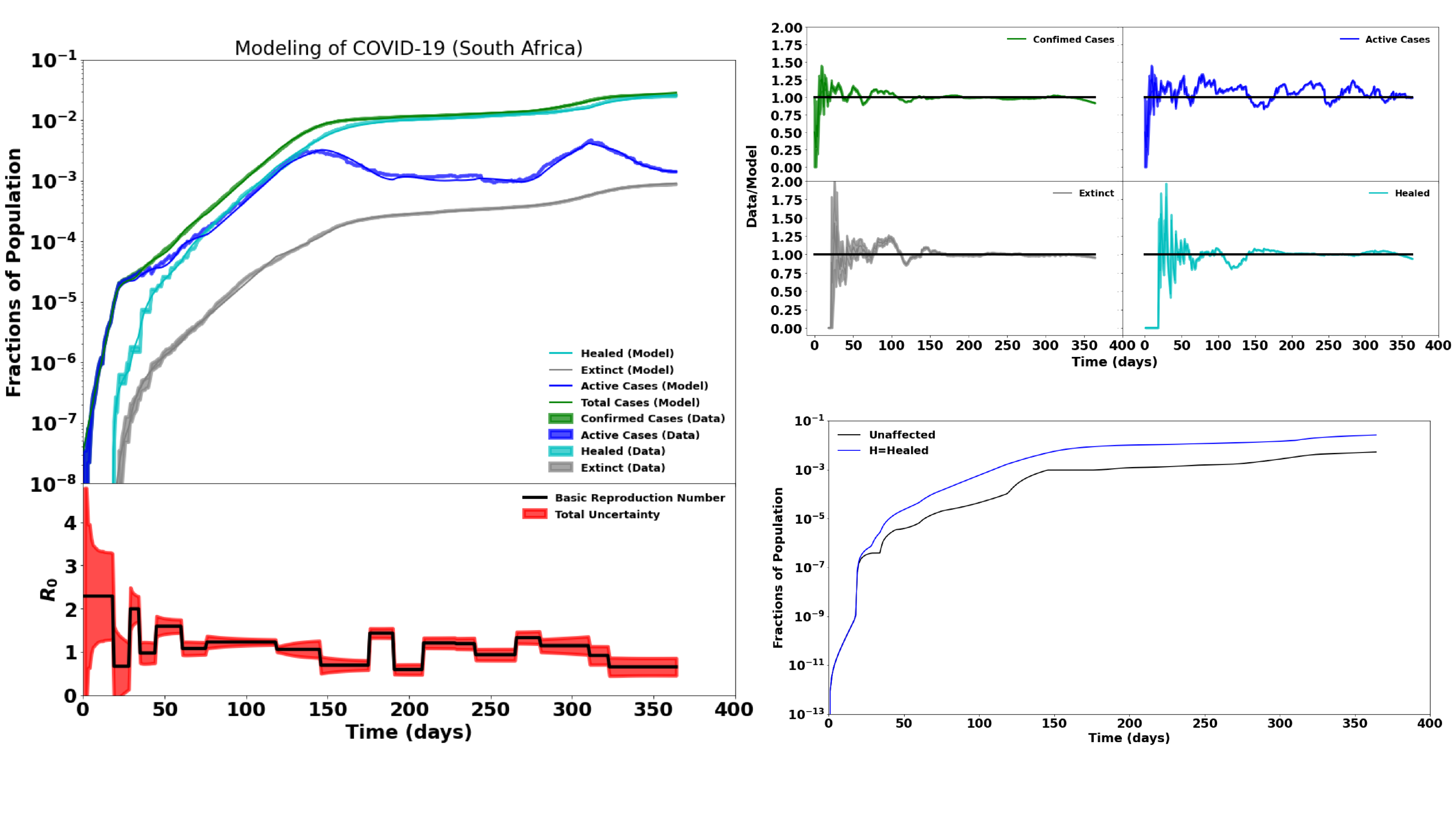}
  	\caption{COVID-19 data and model of South Africa. Active, recovered, death and total cases are shown in the top-left plot. Day~0 is March 5, 2020. The bottom-left plot shows the time-dependent basic reproduction number. The error bands are statistical in the top plot. The bottom-left plot error band includes systematic uncertainty from the infected but unaffected population not counted in the data. The goodness-of-fit of the data modeling is shown as the ratio of the data over the model in the top-right plot. The uncertainty ban contains the statistical uncertainty in the data and the systematic uncertainty on the modeling. The model prediction of the recovered population is shown in the bottom-right plot; also shown, is the undiagnosed fraction of the people that were infected and recovered without symptoms. This fraction, called the unaffected cases, is not measured or included in the data.}
   	\label{fig:SouthAfrica}
  \end{center}
\end{figure}

We observe from the bottom-right panel of Figure~\ref{fig:SouthAfrica} that at the emergence of COVID-19 outbreak in South Africa, the basic reproduction number was $R_0 = 2.25$. Up to March 29, 2020, infections grew exponentially. Lockdown level~5 was effective and the infection rates decreased as shown in Figure~\ref{fig:SouthAfrica}. 
%, which shows that the disease was persistence in the population. 
%However, from March 29, we see a decline in the basic reproduction number because of the 
Throughout the pandemic there has been the implementation of COVID-19 control measures such as social distancing, mask wearing and hand sanitizing. %, local and international travel restrictions and other lockdown regulations. 
%We note that whenever the lockdown regulations were more relaxed, the basic reproduction number increased compared to when the strictest levels were implemented. 
%Validation of the modeling is shown as a ratio of data to the model results in Figure~\ref{fig:SouthAfrica}, top right.
%A feature of the SIDARTHE model is the extraction of the unaffected population. This is the population that is infected, asymptomatic and not diagnosed. 
Until March 29, 2020, in terms of the SIDARTHE model, the unaffected and recovered fraction of population were comparable as seen in Figure~\ref{fig:SouthAfrica}, bottom right. 
There is a correlation between the basic reproduction number and the implementation of the various lockdown levels.

Around the start of lockdown level~1, from September 15 to about beginning of October, $R_0 > 1$. From the end of October to beginning of November, $R_0 \sim 1$.  %This was because of the rising number of daily new cases and deaths. 
%This signaled the coming 
The second wave then started around November 8 with a slow rise to the beginning of December 1. On December 3, the region of Nelson Mandela Bay was put under stricter lockdown rules, including a ban in alcohol sales and tighter restrictions on gatherings---such as funerals and prohibition of the reopening of schools.  Nonetheless, the spread of the virus %became uncontrollable and 
increased sharply.
This second wave was accompanied by the emergence of the new variant, 501Y.V2, in South Africa~\cite{Tegally2021}.

On December 29, the whole country went back to an "adjusted" alert level~3.  The number of daily confirmed cases kept rising until January 11, 2021---around when the second wave peaked. Thereafter, the number of cases started to decrease and $R_0 \sim 1$.
%The second wave impacted rural areas because of festive season travels; the first wave did not impact severely most rural areas.

\subsection{Analysis of COVID-19 data of Ghana} 
\begin{figure}[!h]
 \begin{center}
 	\includegraphics[width=\textwidth]{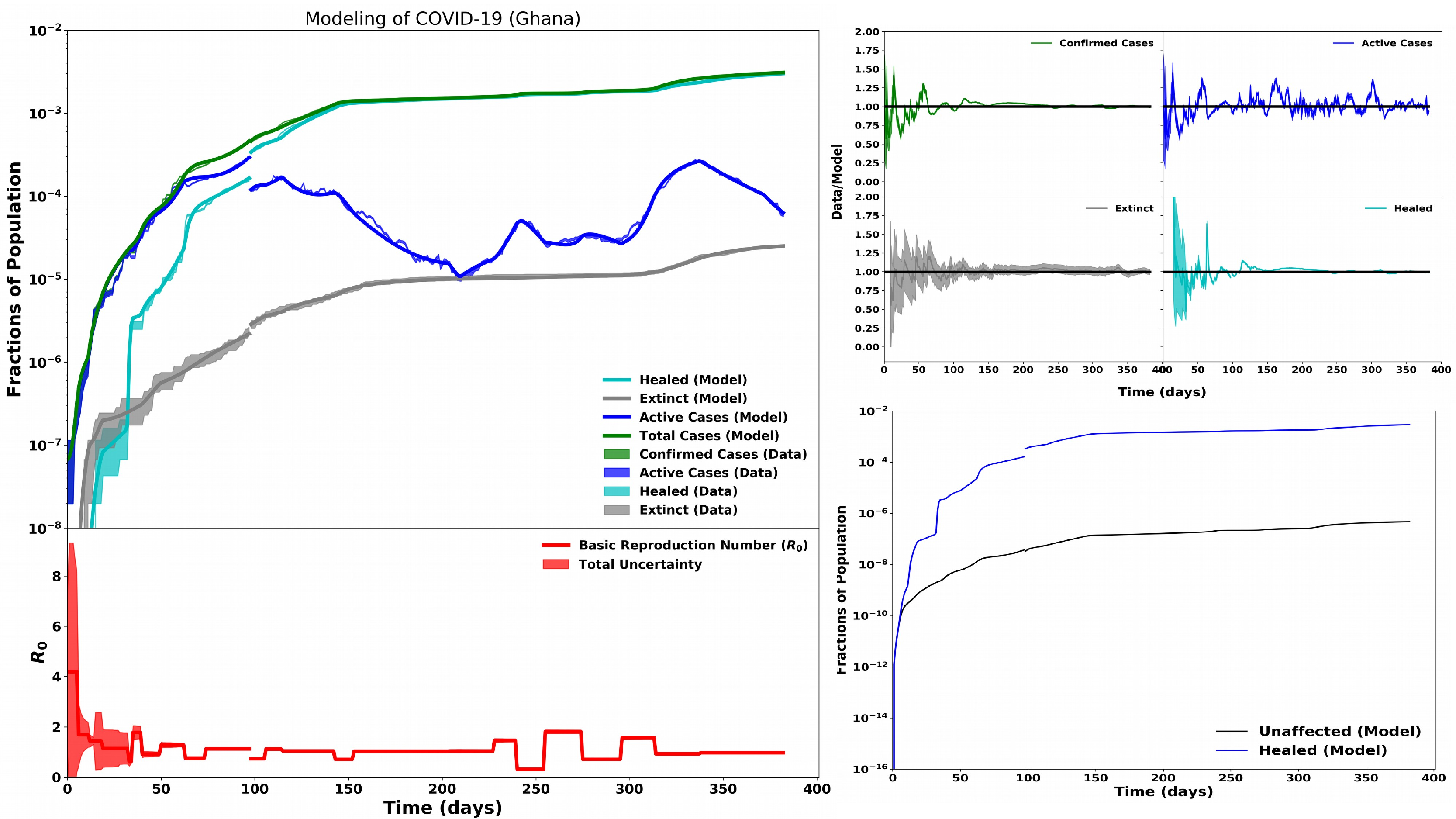}
  	\caption{COVID-19 data and model of Ghana. Active, recovered, death and total cases are shown in the top-left plot. Day~0 is March 12, 2020. The bottom-left plot shows the time-dependent basic reproduction number. The error bands are statistical in the top plot. The bottom-left plot error band includes systematic uncertainty from the infected but unaffected population not counted in the data. The goodness-of-fit of the data modeling is shown as the ratio of the data over the model in the top-right plot. The uncertainty ban contains the statistical uncertainty in the data and the systematic uncertainty on the modeling. The model prediction of the recovered population is shown in the bottom-right plot; also shown, is the undiagnosed fraction of the people that were infected and recovered without symptoms. This fraction, called the unaffected cases, is not measured or included in the data.}
   	\label{fig:Ghana}
  \end{center}
\end{figure}

\noindent The first two imported cases of COVID-19 in Ghana were announced on March 12, 2020. Four days later, a total of six cases were identified. Contact tracing led to 143 suspected cases on March 17 and triggered the deployment of preventive measures~\cite{GhanaRef1, GhanaRef2}. With immediate effect, the government enacted a ban on public gatherings and discouraged airlines to accept travelers from countries with more than 200 cases. The government ordered a series of market disinfection exercises. The first phase was carried out in March while the second and third phases were performed in July and November 2020 respectively. From March 23, the Ghanaian authorities closed borders with the neighboring countries. Furthermore, some local cities were subjected to a partial lockdown until April 19. Local organizations, companies, businesses, and generous citizens supported the containment measures with gifts of locally-designed reusable or medical face masks, veronica buckets, hand sanitizers, and other hygienic items. On May 1, 2020---fifty days since the start of the outbreak in Ghana---there were 2074 confirmed cases, with 200 recoveries and 17 deaths. The top-left plot of Figure~\ref{fig:Ghana} shows the modeling of the Ghanaian data; in the bottom-left plot of Figure~\ref{fig:Ghana}, we see an increase in $R_0$ from 0.91 to 1.28 on day 50, followed by a fall on May 14. 
This reduction in the cases resulted from measures imposed in March and April, 2020. By May, more than 80\% of Ghanaian regions reported their first cases although the total confirmed cases was less than 0.02\% of the population and the country had seen two waves of the pandemic. Meanwhile, the restrictions on bars and restaurants were lifted and this resulted in an increase in $R_0$ to 1.12 at the end of May. By June 9, all the 16 regions had registered their first cases. Since the beginning of the epidemic, the Ghana Health Service has implemented a discharging protocol that declared recovered a suspected COVID-19 patient who tested negative two consecutive times. However, from June 18, there was a change in policy: asymptomatic infected, and symptomatic infected persons whose symptoms were mitigated by treatment, were discharged. The new protocol was adopted to relieve the health system and comply with new directives from the WHO. This led to a rapid increase in the recovered cases, from 4548 to 10074 as shown in Figure~\ref{fig:Ghana}, top-left plot. The sudden increase in the number of recovered cases was not a reflection of the pandemic evolution; therefore, we modeled two datasets. In the first model, we considered the data from March to June 17; in the second modeling, the data from June 18 was used. By the end of July, 2020, the total cases increased from 19388 to 37812. It was the first time the monthly increase fell below a 100\% in Ghana. $R_0$ fluctuated around one from June to the beginning of October, suggesting that the situation in Ghana was stabilizing steadily. However, a surge in new cases during the second half of October increased the $R_0$ from 1.03 to 1.46---this increase could be related to the re-opening of air borders and schools; in November, $R_0$ fell again to 0.31. Two subsequent waves occurred towards the ends of November and December and $R_0$ increased from 0.31 to 1.81 and 0.71 to 1.56 respectively. By February 12, 2021, the total number of cases was 77046 with 7778 active cases, 68713 recovered cases and 555 deaths. The top-right plot of Figure~\ref{fig:Ghana} shows the validation of the modeling of the COVID-19 of Ghana in a ratio of data to model expectations as function of time. We also estimated the fraction of the population that was infected but unaffected; this is shown in bottom-right plot of Figure~\ref{fig:Ghana}.

\subsection{Analysis of COVID-19 data of Kenya}
\noindent The first case was reported on March 12, 2020. It was a young lady who had traveled back to the country from the US via London. The second case was a friend of Patient Zero and the third case was one of the passengers sitting next to Patient Zero on the flight from London to Nairobi. Two weeks after the first case was reported, a cessation of movement from the country’s two largest cities, Nairobi and Mombasa, was introduced to curb the spread of the virus from the two cities which were slowly emerging as hot spots. Other measures taken by the Kenyan government were: 1) the country went into a partial lockdown; 2) both international and local flights were suspended; 3) schools were closed; 4) night curfews were introduced and are still in place; 5) social distancing was required in restaurants and other social places; 6) social gatherings were limited to a maximum of 50 people; and lastly, 7) people were encouraged to work from home where possible between the months of April-August 2020~\cite{kenya2020}. 
\begin{figure}[!h]
 \begin{center}
 	\includegraphics[width=\textwidth]{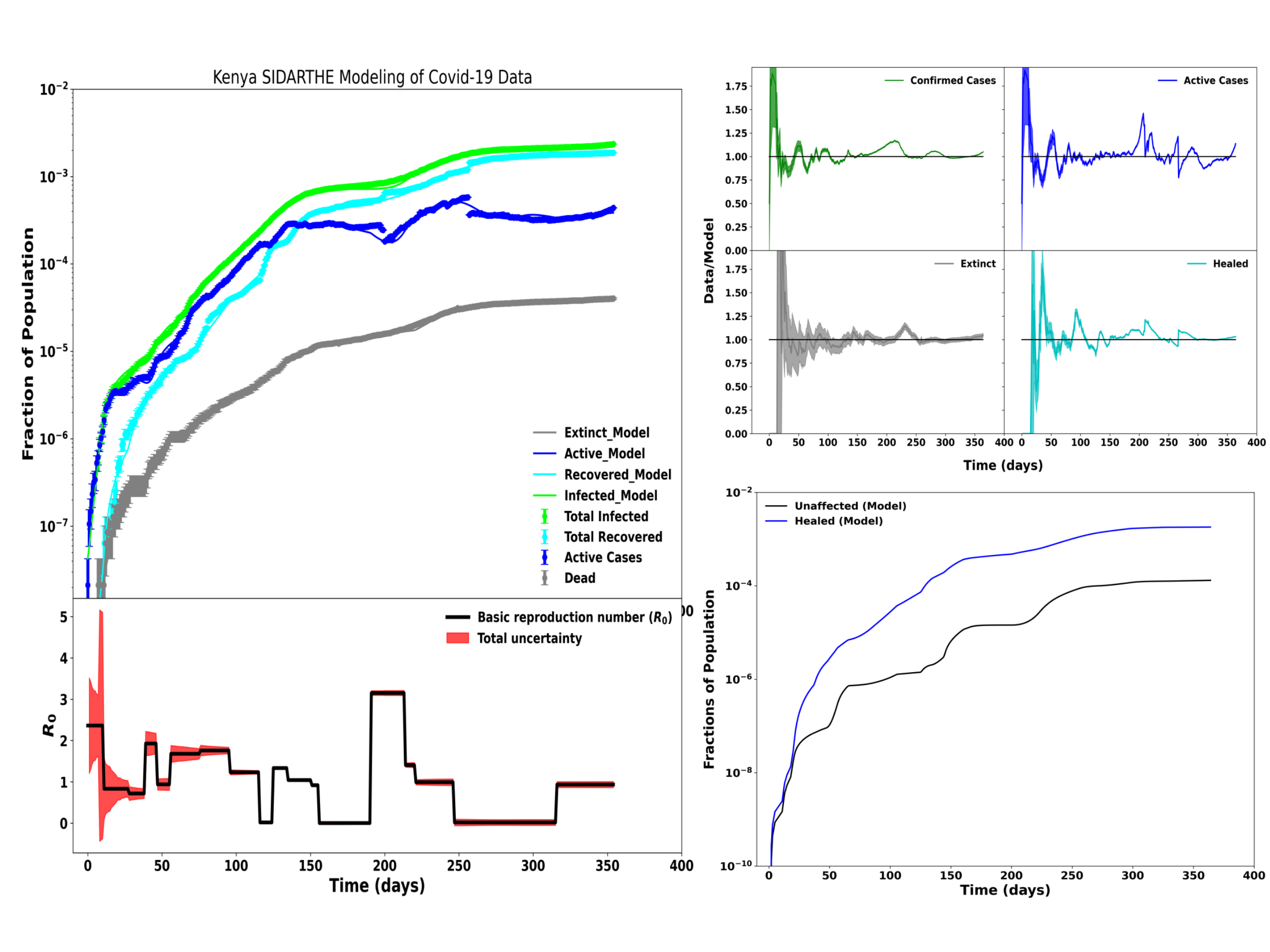}
  	\caption{COVID-19 data and model of Kenya. Active, recovered, death and total cases are shown in the top-left plot. Day~0 is March 12, 2020. The error bars show the statistical errors between the Model and the Kenya C0VID-19 data. The bottom-left plot shows the time-dependent basic reproduction number and the error band includes systematic uncertainty from the infected but unaffected population not included in the data. The top-right plot shows the goodness-of-fit which is given by the ratio of the data over the model. The uncertainty ban contains the statistical uncertainty in the data and the systematic uncertainty on the modeling. The model prediction of the recovered population is shown in the bottom-right plot; also shown, is the undiagnosed fraction of the people that were infected and recovered without symptoms. This fraction, called the unaffected cases, is not measured or included in the data.}
   	\label{fig:Kenya}
  \end{center}
\end{figure}
 As shown in the top-left plots of Figure~\ref{fig:Kenya}, numbers of cases rose steeply in the first month of the pandemic and slowed down for the next three months as the infection rates fell. The flattening of the curve was achieved around the month of July. Five months after the first case was reported, the total recovered cases surpassed the total active cases as the number of infections continued to fall. The fatality rate was at $\sim$1.5\%. The top-right plot of Figure~\ref{fig:Kenya} shows how well the Kenyan data is modeled. An estimate of the unaffected fraction of the Kenyan population is shown in bottom-right plot of Figure~\ref{fig:Kenya}.

At the beginning, $R_{0}\,=\,2.4$ as the pandemic was growing from the first imported cases, and the government had not yet introduced any control measures. However, $R_{0}$ drastically fell after the second week as shown in the bottom-left plot of Figure~\ref{fig:Kenya}. A month after the first case was reported, $R_{0} \leq 1$ and remained below one indicating that the number of new infections was decreasing. This can be attributed to the closure of entry points, i.e. suspension of international flights, which eliminated imported cases. The gradual decline in the $R_{0}$ implied that the control measures introduced were effective in curbing the spread of the virus in those earlier days. However, as shown in Figure~\ref{fig:Kenya}, bottom-left plot, $R_{0}$ periodically rose and fell due to a number of reasons such as the increased testing capacity and relapses in adherence to the control measures.  

Since January 2021, normalcy has resumed; however, people are encouraged to wear masks in public places and observe social distancing. Schools have reopened and international and domestic flights resumed.

\subsection{Analysis of COVID-19 data of Madagascar}
\begin{figure}[!h]
 \begin{center}
 	\includegraphics[width=\textwidth]{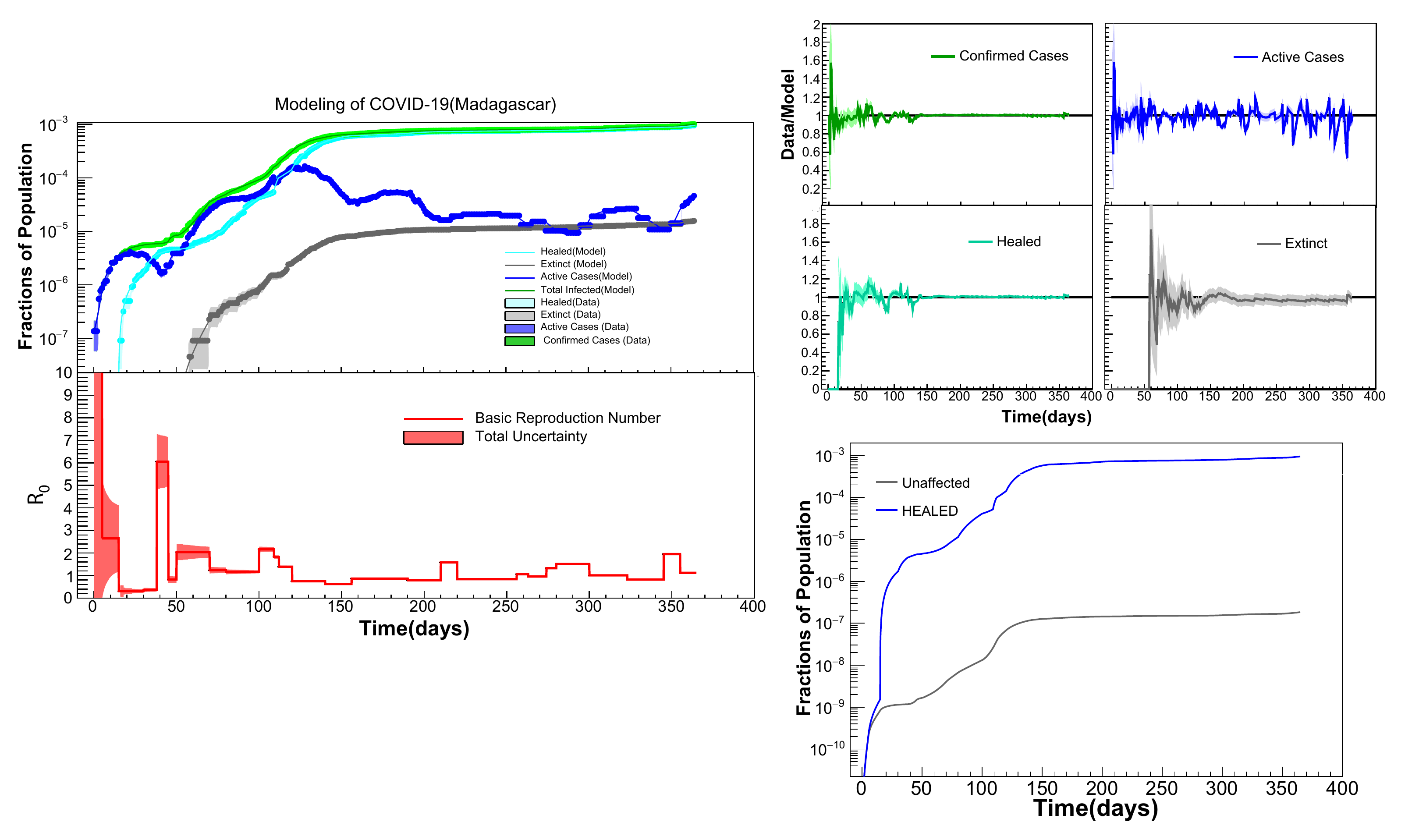}
  	\caption{COVID-19 data and model of Madagascar. Active, recovered, death and total cases are shown in the top-left plot. Day~0 is March 13, 2020. The bottom-left plot shows the time-dependent basic reproduction number. The error bands are statistical in the top plot. The bottom-left plot error band includes systematic uncertainty from the infected but unaffected population not counted in the data. The goodness-of-fit of the data modeling is shown as the ratio of the data over the model in the top-right plot. The uncertainty ban contains the statistical uncertainty in the data and the systematic uncertainty on the modeling. The model prediction of the recovered population is shown in the bottom-right plot; also shown, is the undiagnosed fraction of the people that were infected and recovered without symptoms. This fraction, called the unaffected cases, is not measured or included in the data.}
   	\label{fig:Madagascar}
  \end{center}
\end{figure}

\noindent On March 13, 2020, 117 days after the first case of COVID-19 appeared in Wuhan, China, Madagascar identified its first three confirmed cases in the capital city of Antananarivo. From that date until February 23, 2021, 119608 tests have been conducted resulting in 19831 confirmed cases. At the time of writing, two hundred and thirty eight cases were still active and under treatment, while 19296 recovered with a death toll of 297~\cite{MadagascarRef1, MadagascarRef2}. The modeling of the Malagasy COVID-19 and the $R_{0}$ are shown in the right plots of Figure~\ref{fig:Madagascar}, the validation of the modeling is in top-right plot and an estimate of the unaffected population in the bottom-right plot of Figure~\ref{fig:Madagascar}.
The local authorities dealt with the crisis in four major steps. First, since the economy of Madagascar relies heavily on tourism industries, the government decided not to close the borders. Approximately three weeks after the first cases, confirmed positive cases rose to fifty seven. The second step was initiated because of the increase in cases and a better understanding of the disease based on WHO recommendations; the government initiated a lockdown of the country by suspending all international flights for thirty days during the month of April 2020.  The majority of the cases were identified in the two major cities of the country~\cite{MadagascarRef3}. At fifteen-day increments, local lockdowns were enforced within these two cities and all regional flights were suspended as well.  Masks were required and enforced heavily by local forces. Drastic sanctions against non-mask wearers were put in place. At the same time, a newly developed herbal tea called Covidorganics was distributed for free. These measures mitigated the transmission of the disease; by the end of the second fifteen-day incremental lockdown, the number of cases was 132, with only 36 active cases under treatment. Madagascar started free testing and put in place several new sites to conduct data collection of infected peoples. Meanwhile, the fragile economy started to crumble under the heavy lockdown measures; therefore, sanctions were loosened and population mobility resumed; as a result, the number COVID-19 cases increased. By the end of May 2020, 771 cases were confirmed, with the first recorded death. In the following months, six additional deaths, caused by COVID-19, occurred. The third step was a full lockdown of the country, which was difficult to enforce since local economy in micro and small enterprises is at the core of the Malagasy culture~\cite{MadagascarRef4}. The confirmed cases then rose to 10890 by the end of July 2020 with 106 deaths. The government provided a stimulus package and social programs to support the population, but at the cost of large gatherings to receive the government assistance. The number of cases were also  affected by a large repatriation of Malagasy nationals from Europe. When the fourth step started in August 2020, the number of new cases kept decreasing from 1522 in September to 380 in December 2020~\cite{MadagascarRef1, MadagascarRef2}.  All international flights were still suspended, local lockdowns were  lifted and case reporting changed from daily to weekly. At the time of writing, the authorities were evaluating the evolution of COVID-19 abroad, and if the situation continued to improve, international flights might resume. 

\subsection{Analysis of COVID-19 data of Cameroon} 
\begin{figure}[!h]
 \begin{center}
 	\includegraphics[width=\textwidth]{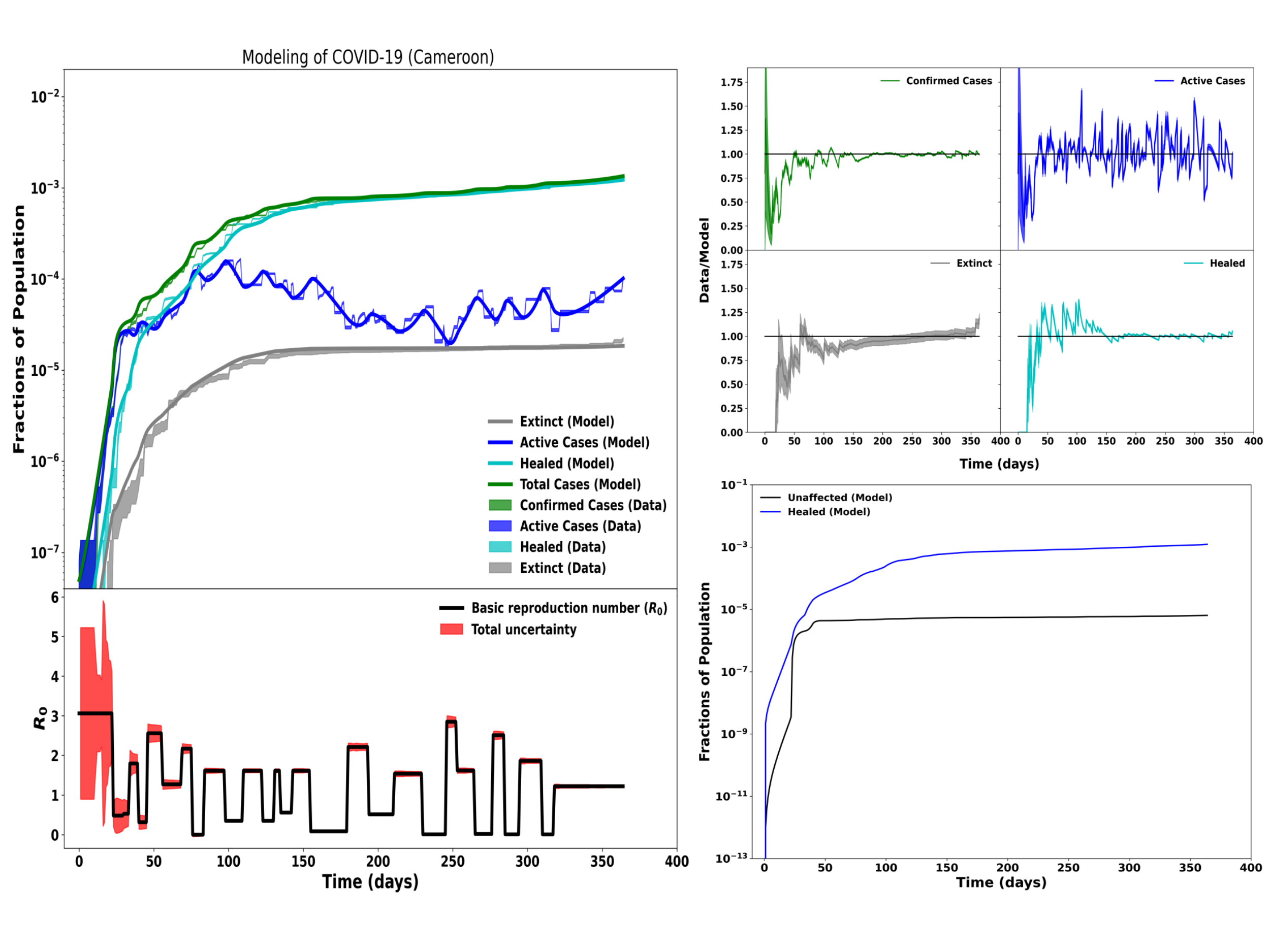}
  	\caption{COVID-19 data and model of Cameroon. Active, recovered, death and total cases are shown in the top-left plot. Day~0 is March 6, 2020. The bottom-left plot shows the time-dependent basic reproduction number. The error bands are statistical in the top plot. The bottom-left plot error band includes systematic uncertainty from the infected but unaffected population not counted in the data. The goodness-of-fit of the data modeling is shown as the ratio of the data over the model in the top-right plot. The uncertainty ban contains the statistical uncertainty in the data and the systematic uncertainty on the modeling. The model prediction of the recovered population is shown in the bottom-right plot; also shown, is the undiagnosed fraction of the people that were infected and recovered without symptoms. This fraction, called the unaffected cases, is not measured or included in the data.}
   	\label{fig:Cameroon}
  \end{center}
\end{figure}

\noindent The first case of COVID-19 in Cameroon was announced on March 6, 2020 and the government implemented containment measures described in Ref.~\cite{CameroonRef1}. The left plots of Figure~\ref{fig:Cameroon} show the modeling of the COVID-19 data of Cameroon and the estimate of the basic reproduction number. In top-right plot of Figure~\ref{fig:Cameroon}, we show how well the data is modeled. The fraction of the population that is infected but unaffected by COVID-19 is shown in the bottom-right plot of Figure~\ref{fig:Cameroon}.
The government advised the public not to panic, but to show discipline, solidarity and a sense of responsibility. These measures helped reduce the rate of pandemic evolution until the government eased the measures on October 15, 2020. From then, there was a sharp rise in the number of new COVID-19 cases as shown in Figure~\ref{fig:Cameroon}. 

The government increased awareness campaign at the community level, through the media and social networks~\cite{CameroonRef3}. Following the 2020 African Nations Championship soccer tournament, hosted in Cameroon, there was a resurgence of COVID-19 cases and hospitalization rates increased from 0.5\% to 5\%~\cite{CameroonRef4}. 

\subsection{Analysis of COVID-19 data of Mozambique}
\noindent References~\cite{mozambique1, mozambique2, assamagan2020study} describe the first few months of the COVID-19 evolution in Mozambique since the first case was identified on March 22, 2020. On May 12, they suspended international flights until May 30, except for humanitarian, cargo or state flights. However, universities and schools reopened with social distancing, hand washing and mask wearing required. Masks were mandatory in public spaces and private businesses, and enforced in public transportation and hospitals; bars and museums remained closed. We show in the top-left plot of  Figure~\ref{fig:Mozambique} the modeling of the COVID-19 data of Mozambique. In the top-right plot of Figure~\ref{fig:Mozambique}, we compare the modeling results to the data, and an estimate of the unaffected population is presented in the bottom-right plot.

The council of ministers approved a decree, “Situation of Public Disaster and Activated Red Alert”, which came into effect on September 7~\cite{mozambique3}; non-compliance was punishable with fines up to five minimum wages. Among the measures imposed in the decree, prevention and containment measures stood out: usage of masks, frequent hand washing, social distancing, usage cough etiquette and no sharing personal utensils. Passenger transport flights to certain countries resumed, on a reciprocal basis. On October 1, 2020, schools resumed for the twelfth grade education, with contingency plans. For public services and religious celebrations, the number of participants had to be $\leq 50$\% the venue capacity, with a cap at a hundred and fifty people. There were also restrictions on public transports~\cite{mozambique4}.

There were many violations of the containment measures during the festive season, in December 2020, because of  parties, gatherings and concerts with greater than capacity, where social distancing was not observed; this led to an increase in the confirmed cases as shown in Figure~\ref{fig:Mozambique}, top plot. In response, the government implemented a new decree in January 2021, namely COVID-19 screening test before entering or leaving the country; stay at home order for people with symptoms of fever and flu; restriction in the opening hours of all commercial activities, to 6pm; closing of all booths selling alcoholic beverages; suspension cultural activities. These measures entered into force on January 15 for 21 days~\cite{mozambique5}. However, the number of cases continued to grow, and by the end of January 2021, the number of active cases doubled---until December 31, 2020 they had a total of 18642 confirmed cases and on January 31, 2021, it was 38654. Additional and more restrictive measures were imposed in early February 2021: no religious ceremonies or public services; interdiction of social and private gatherings, except weddings with no more than twenty people; prohibition of sales of alcoholic beverages; closure of schools; limitation to one person-visit per month to penitentiary establishments~\cite{mozambique6}. 

We see good agreement between the modeling and the data for the dead, recovered and active cases of the population, as shown in Figures~\ref{fig:Mozambique}. As a result, the total cases are also well modeled. The estimated $R_0$ remains below two for the entire period shown in bottom-left plot of Figure~\ref{fig:Mozambique}. The $R_0$ for Mozambique fluctuated over time: between days 29 to 39, 54 to 62, 145 to 154, 162 to 166, it dropped significantly. In days 3 to 29, 75 to 112, 116 to 135 it stays slightly above one.

\subsection{Analysis of COVID-19 data of Rwanda}
\noindent We described the first few months of the COVID-19 pandemic---since the first case was detected on March 14, 2020---in Ref.~\cite{assamagan2020study}. Testing of symptomatic cases started right away before the first case was identified, especially those coming from outside~\cite{RwandaRef1}. Contact tracing and testing of asymptomatic cases started after April 7; the Rwandan government used contact tracing by testing all individuals ---whether they show symptoms or not---who came in contact with an infected person~\cite{RwandaRef1}.

The data and model results are shown in the top-left plot of Figure~\ref{fig:Rwanda}; the estimated $R_0$ is shown in the bottom-left plot. In the top-right plot, we present the validation of the modeling; the fraction of the Rwandan population unaffected by COVID-19 is shown in the bottom-right plot of Figure~\ref{fig:Rwanda}. On January 18, 2021, Rwanda imposed a further lockdown in the city of Kigali because of a resurgence of COVID-19 cases with increased transmission and death rates. The government informed the public that all movements outside residential homes required authorization except for essential services. Movements between districts/provinces were prohibited except for tourism and essential services; however, tourists were required to show a negative COVID-19 test certificate.  Public transport was also prohibited and all employees worked from home, except those who provided essential services. Places of worship and learning institutions remained closed until further notice; night curfew started at 18:00 until 4:00 local time~\cite{RwandaRef1}.

\subsection{Analysis of COVID-19 data of Togo}
\noindent In Ref.~\cite{assamagan2020study}, we summarized the first three months of the pandemic; the first case was confirmed on March 6, 2020~\cite{RepubliqueTogolaise, agbokou2020investigation}. From mid-June until October,  examination classes resumed for students. Also, on August 1, 2020, international flights restarted. From mid-September, most cases were detected in travelers~\cite{agbokou2020investigation}.

The modeling, validation and the unaffected fraction of the Togolese population are shown in Figure~\ref{fig:Togo}. In November 2020, both primary and high schools resumed with all the measures---such as hand washing and wearing of masks---in effect. At the time of writing, only a few infections were detected in pupils~\cite{RepubliqueTogolaise}. During the 2020 end of the year holiday season, curfew was imposed from 10:00pm to 6:00am. Despite the curfew and all the preventive measures, the number of cases increased substantially in the northern region of the country from January 15, 2021, as shown in Figure~\ref{fig:Togo}, top-left plot. This was because of violations of the preventive and social distancing measures~\cite{sadio2021assessment, agbokou2020investigation}. As a result, from January 17, 2020, the government implemented a new lockdown in the regions with increased infection rates, for a period of three weeks. A curfew from 8:00pm to 5:00am was also imposed.

\subsection{Analysis of COVID-19 data of Zambia} 
\noindent The first three months of COVID-19 are described in Refs.~\cite{assamagan2020study, ZNPHI} since the first two cases of COVID-19 on March 18, 2020. The top-left plot of Figure~\ref{fig:Zambia} shows the COVID-19 data of Zambia and its SIDARTHE modeling. The validation of the modeling for Zambia, and the unaffected fraction of the Zambian population are shown in right plots of Figure~\ref{fig:Zambia}.

As shown the bottom-left plot of Figure~\ref{fig:Zambia}, the $R_0$ value around day 90 reflected a gradual relaxation of physical distancing measures in May and June 2020.  However, the $R_0$ remained below 2 for all the dates, apart from around day 130; this increase is due to an increase in testing. Some reports suggested that the number of COVID-19 infections is likely to be higher than the confirmed case counts because numerous infected people have moderate or no symptoms, and limitations exist concerning testing capacity and surveillance systems in Zambia~\cite{MULENGA2021}. 

\section{Discussion}
\label{sec:disc}
\noindent In our earlier studies, reported in Ref.~\cite{assamagan2020study}, we concentrated on the first three months of COVID-19 data. The studies presented in this paper are an extension of the earlier ones with more countries included and data up to twelve months. We estimated the basic reproduction numbers by simultaneous fits of the SIDARTHE model to the data of infected, recovered and dead cases. The modeling followed the major changes in the data patterns that could be linked to the pandemic evolution and the response measures applied.  We note cyclic rises and falls in the $R_0$ as a function of time. Where no correlations between the $R_0$ and the control measures can be made, other less obvious effects---such as difficulties to adhere to government directives or porous regional borders---might account for the observed patterns.

In many of the cases studied, the numbers of tests done were comparatively small given the population sizes. We estimated the statistical uncertainties in the data and propagated these to the uncertainties on $R_0$. At the start of the pandemic when the data was small, estimates of $R_0$ had large uncertainties. These uncertainties reduced as the number of tests increased, thus the statistical uncertainty on $R_0$ diminished. The systematic uncertainties have different sources such as the methods of testing and data collections, and modeling errors. In the current studies, we only estimated one source of systematic uncertainty in the modeling: the SIDARTHE model of the healed population includes the fraction of the infected people that are unaffected by the virus. However, the data does not measure this fraction. We estimated this fraction as a systematic uncertainty in the modeling and we combined it in quadrature with the statistical uncertainty to obtain the total uncertainties on the $R_0$ estimates. The experimental systematic uncertainties require further inputs from the methods of testing and data collections, and these are beyond the scope of the current studies.

%Inserted by Simon
It is important to consider why the data does not show a similar severity of the disease in Africa as compared to Europe, America and some countries in South America~\cite{Lalaoui2020}. This is despite earlier studies predicting especially dire scenarios for the course of the pandemic for Africa. Many hypotheses for this have been presented, and we comment on this in the light of our own analyses. Consider a snapshot from  Our World in Data, Statistics and Research, Coronavirus (COVID-19) Cases, expressed as the 7-day average of daily new confirmed COVID-19 cases (deaths) per million people~\cite{Orce2021}. On March 7, 2021, there were the following results: Africa 7.12 (0.21), South Africa 18.59 (1.65), Europe 187.05 (4.17), United States 179.85 (5.09). This puts the clinical prevalence for Africa at 51 (44) times lower than the average of the United states and Europe, and for South Africa those numbers are 20 (5.6). The numbers for South Africa are singled out as it is the country which has had the highest clinical prevalence in Africa.

A recent study of seroprevalence of anti-SARS-CoV-2 antibodies among blood donors in several provinces of South Africa was conducted in January 2021. At the time of writing, this research was a preprint~\cite{Sykes2021}. It studied 4858 donors and found that the weighted estimates of prevalence by province, in the age range of 15-69 year old, vary from 32 to 63\% compared to clinically-confirmed cases in mid-January 2021 of 2.2 to 2.8\%. 
%"weighted net estimates of prevalence, in the core age range 15–69, by province (compared with official clinically-confirmed COVID-19 case rates in mid-January 2021) are: EC-63\%(2.8\%), NC-32\%(2.2\%), FS-46\%(2.4\%), and ZN-52\%(2.4\%)". 
The sampled group would be a mixture of the susceptible and healed groups in the SIDARTHE model. This indicates that the ratio of total infected to clinically-confirmed cases would be at least an order of magnitude. This study would need to be validated and also taken further. However, even if the data in Africa are considerably under-representative of the true prevalence, we would still conclude the incidence of the disease in Africa has not been as severe as in the two wealthy regions considered here.

The reasons for this large disparity which have been presented so far come in two categories. 
Either the situation in Africa is dramatically under-reported, or, indeed, the progression of the disease is less severe in Africa. Included in the former category are some or all of the following reasons: cases going unreported by not being presented at hospitals or clinics, insufficient testing facilities and national testing program, and insufficient systems for contact tracing.  Included in the latter category have been the following reasons: the African national government systematic planned interventions have been effective, the population of Africa is relatively young (lower mortality), the climate is on average warmer (outdoor lifestyles and or lower infectiousness), cross-immunity conferred by the some of the higher disease burden of Africa, benefits derived from other vaccinations (BCG for tuberculosis), the use of antimalarial drugs and the genetic polymorphism of the angiotens in-converting enzyme 2 receptor~\cite{Lalaoui2020}. 

We also note that in our discussion of the trajectory of the disease, we could correlate changes of the basic reproduction number to government planned interventions, in terms of enforced social distancing, encouraging safe social behavior and restricting travel, externally and and internally. These correlations argue that African governments indeed acted to influence the progression of these disease, and that these actions had an observable effect. Also, in South Africa, there is a well-established infrastructure in cities and rural areas, at hospitals and clinics, with respect to treatment of HIV, tuberculosis and other similar diseases. This means that contact tracing is already established. This infrastructure was re-purposed for use in fighting the corona virus pandemic. Similar situations pertain in other African countries. Although there is some under-reporting of the course of the COVID-19 disease in Africa, the last two observations make it unlikely that this is the only reason that the case load is lower. Accordingly, we must consider that Africa has experienced a less severe form of the pandemic, and consider the reasons why this could be so.

\section{Conclusions}
\label{sec:conc}
\noindent We analyzed twelve months of COVID-19 data of Cameroon, Ghana, Kenya, Madagascar, Mozambique, Rwanda, South Africa, Togo and Zambia. For each country, we estimated the time-dependent basic reproduction number, $R_0$. At the onset of the pandemic, $R_0$ was above one in all the cases studied. Over time, the basic reproduction numbers followed the fluctuation patters reflected in the data. The fluctuations were correlated with the control measures imposed and the emergence of new cases. Approximately twelve months since the first cases were detected in the countries studied, all the $R_0$ were about or below one, suggesting that the pandemic had slowed in these countries. However, because the virus may mutate and new waves are likely to occur, we suggest to maintain the control measures until enough vaccines have been administered to reach herd immunity~\cite{VaccineImpactUS}. Our studies also estimated the fractions of the population that were infected and not diagnosed but recovered without symptoms; in general, we find that these fractions are between $1-10$\% of the recovered cases. The modeling of vaccination campaigns and impact of SARS-CoV-2 variants are beyond the scope of this work~\cite{SidartheVaccine}.

\section*{Acknowledgements}
\noindent Toivo S. Mabote would like to thank Professor Doutor Cl\'audio Mois\'es Paulo (Universidade Eduardo Mondlane) for academic advice and mentorship. We acknowledge support and mentorship from the African School of Fundamental Physics and Applications. We received no financial support for this work.

\noindent 

\bibliography{mybibfile}

\newpage

\section*{Supplementary Material}

\renewcommand{\thefigure}{SM1}
\begin{figure}[!h]
 \begin{center}
 	\includegraphics[width=\textwidth]{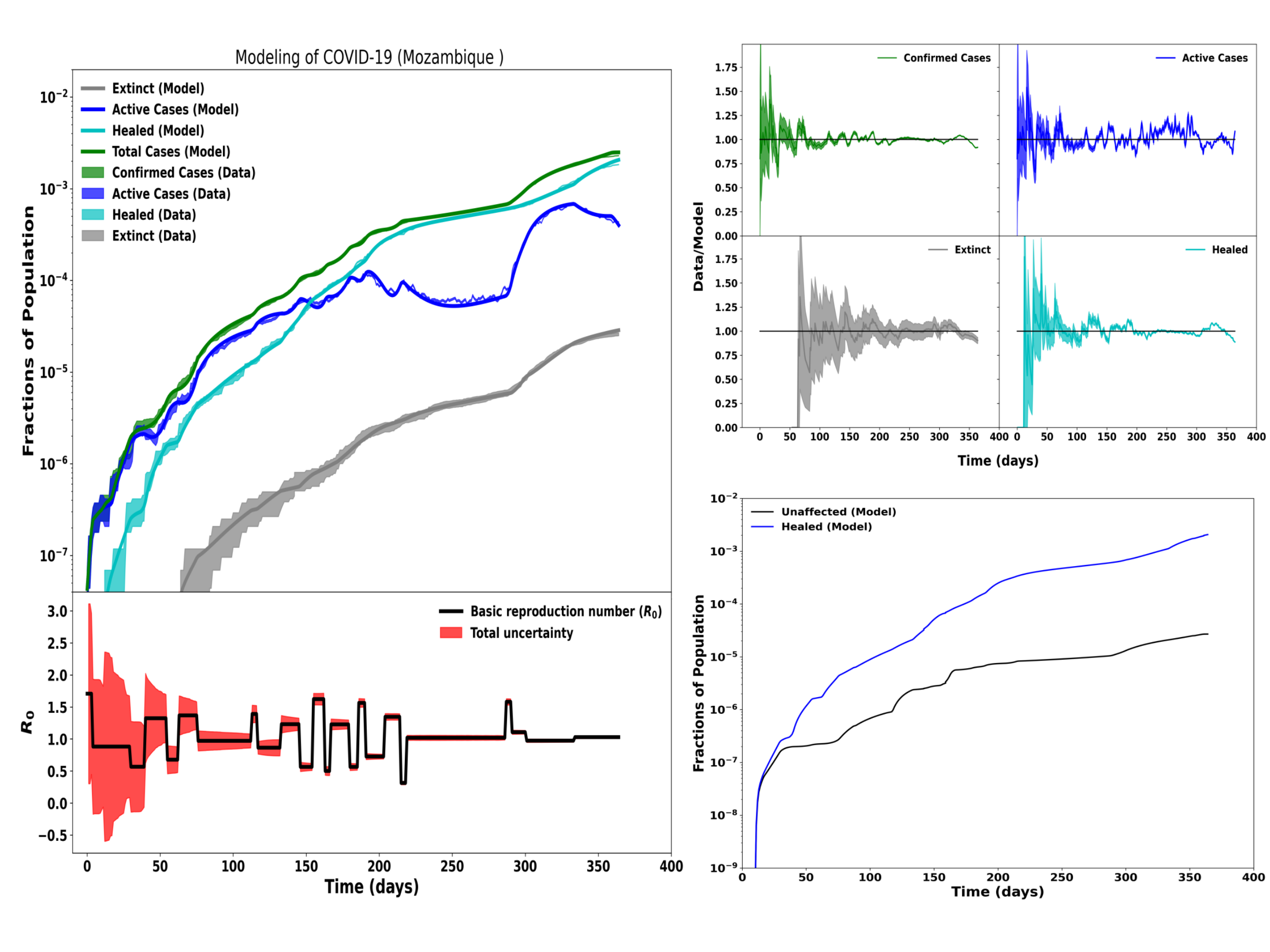}
  	\caption{COVID-19 data and model of Mozambique. Active, recovered, death and total cases are shown in the top-left plot. Day~0 is March 22, 2020. The bottom-left plot shows the time-dependent basic reproduction number. The error bands are statistical in the top plot. The bottom-left plot error band includes systematic uncertainty from the infected but unaffected population not counted in the data. The goodness-of-fit of the data modeling is shown as the ratio of the data over the model in the top-right plot. The uncertainty ban contains the statistical uncertainty in the data and the systematic uncertainty on the modeling. The model prediction of the recovered population is shown in the bottom-right plot; also shown, is the undiagnosed fraction of the people that were infected and recovered without symptoms. This fraction, called the unaffected cases, is not measured or included in the data.}
   	\label{fig:Mozambique}
  \end{center}
\end{figure}
\renewcommand{\thefigure}{SM2}
\begin{figure}[!h]
 \begin{center}
 	\includegraphics[width=\textwidth]{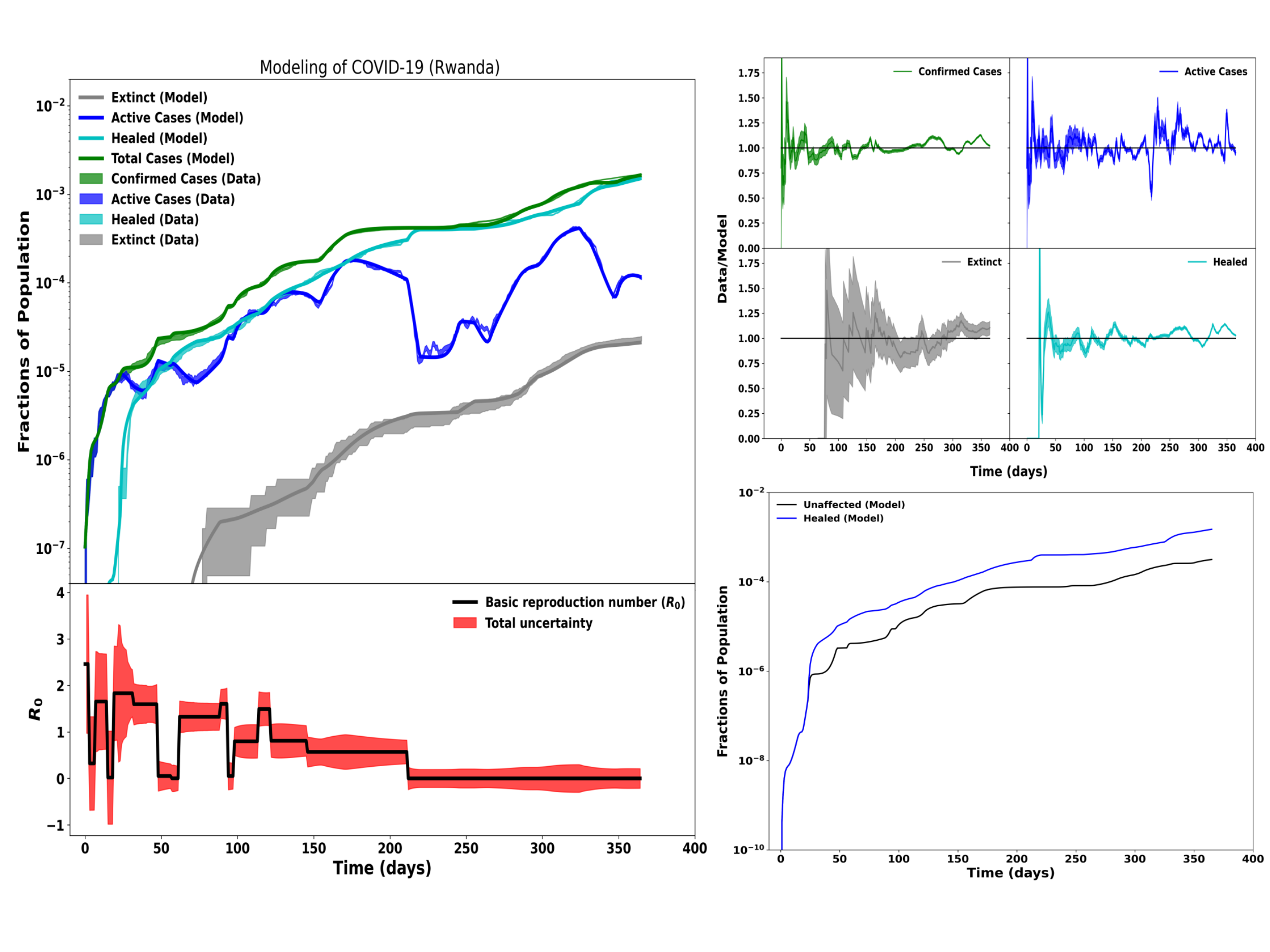}
  	\caption{COVID-19 data and model of Rwanda. Active, recovered, death and total cases are shown in the top-left plot. Day~0 is March 14, 2020. The bottom-left plot shows the time-dependent basic reproduction number. The error bands are statistical in the top plot. The bottom-left plot error band includes systematic uncertainty from the infected but unaffected population not counted in the data. The goodness-of-fit of the data modeling is shown as the ratio of the data over the model in the top-right plot. The uncertainty ban contains the statistical uncertainty in the data and the systematic uncertainty on the modeling. The model prediction of the recovered population is shown in the bottom-right plot; also shown, is the undiagnosed fraction of the people that were infected and recovered without symptoms. This fraction, called the unaffected cases, is not measured or included in the data.}
   	\label{fig:Rwanda}
  \end{center}
\end{figure}
\renewcommand{\thefigure}{SM3}
\begin{figure}[!h]
 \begin{center}
 	\includegraphics[width=\textwidth]{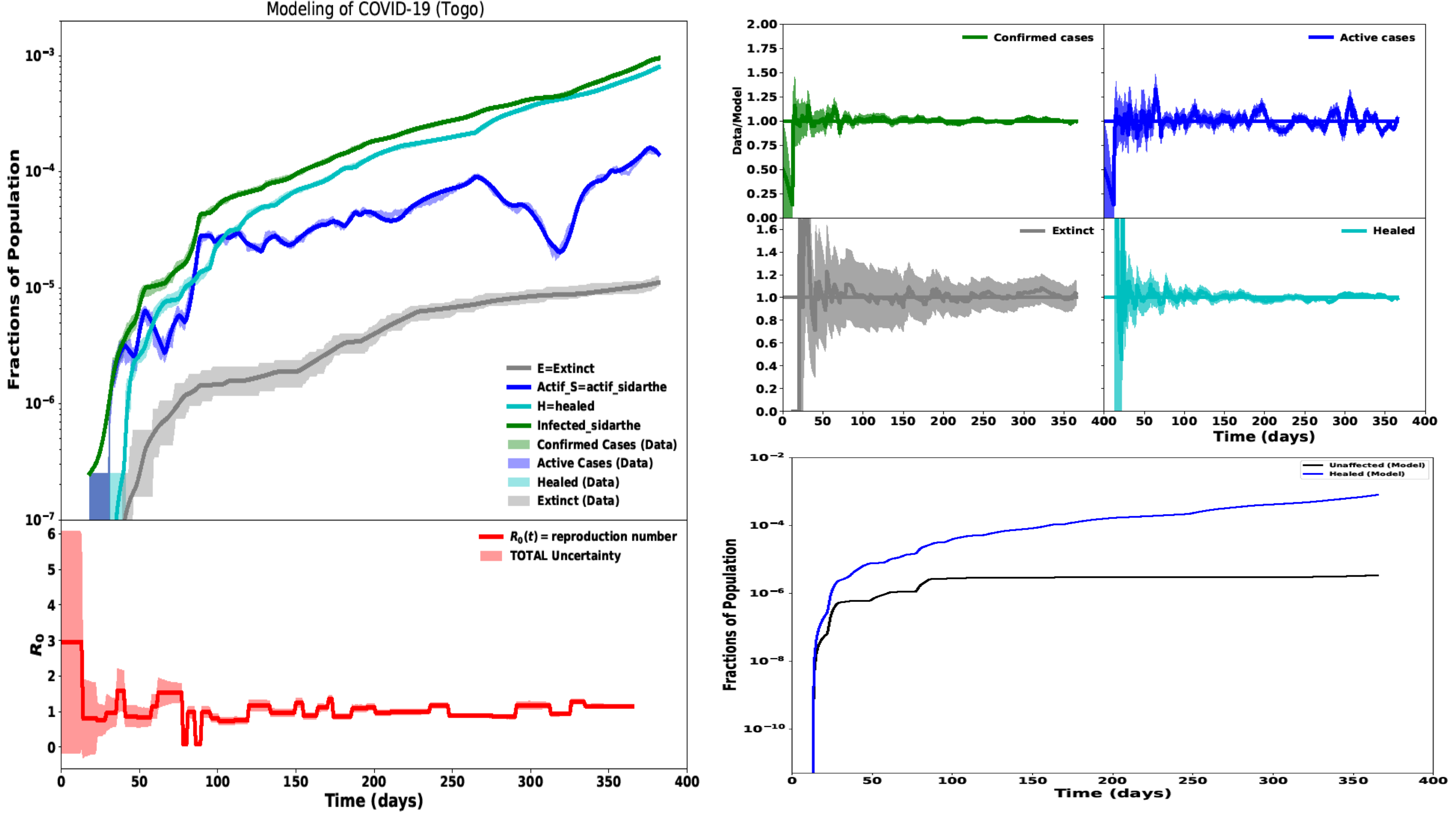}
  	\caption{COVID-19 data and model of Togo. Active, recovered, death and total cases are shown in the top-left plot. Day~0 is March 6, 2020. The bottom-left plot shows the time-dependent basic reproduction number. The error bands are statistical in the top plot. The bottom-left plot error band includes systematic uncertainty from the infected but unaffected population not counted in the data. The goodness-of-fit of the data modeling is shown as the ratio of the data over the model in the top-right plot. The uncertainty ban contains the statistical uncertainty in the data and the systematic uncertainty on the modeling. The model prediction of the recovered population is shown in the bottom-right plot; also shown, is the undiagnosed fraction of the people that were infected and recovered without symptoms. This fraction, called the unaffected cases, is not measured or included in the data.}
   	\label{fig:Togo}
  \end{center}
\end{figure}
\renewcommand{\thefigure}{SM4}
\begin{figure}[!h]
 \begin{center}
 	\includegraphics[width=\textwidth]{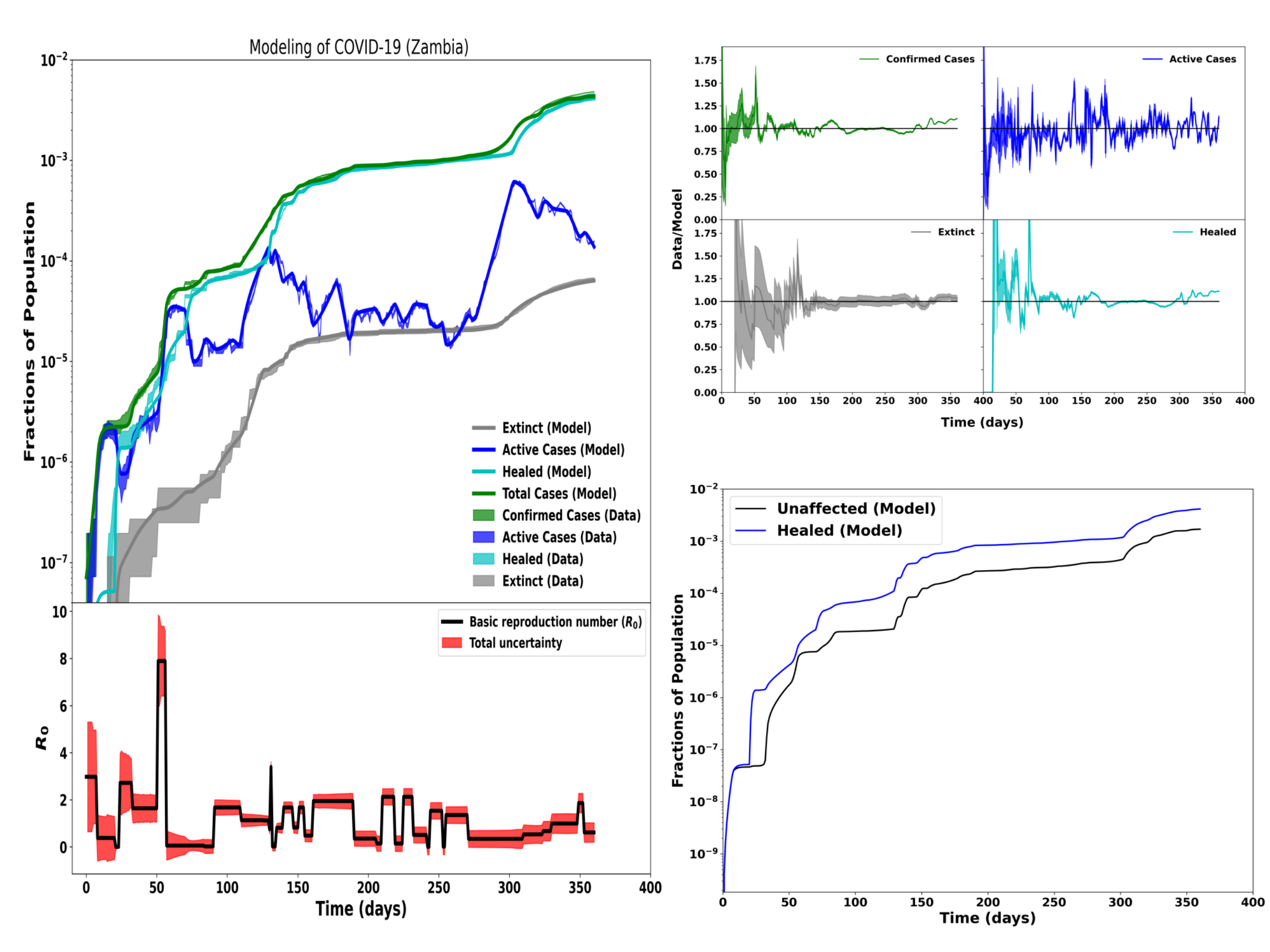}
  	\caption{COVID-19 data and model of Zambia. Active, recovered, death and total cases are shown in the top-left plot. Day~0 is March 18, 2020. The bottom-left plot shows the time-dependent basic reproduction number. The error bands are statistical in the top plot. The bottom-left plot error band includes systematic uncertainty from the infected but unaffected population not counted in the data. The goodness-of-fit of the data modeling is shown as the ratio of the data over the model in the top-right plot. The uncertainty ban contains the statistical uncertainty in the data and the systematic uncertainty on the modeling. The model prediction of the recovered population is shown in the bottom-right plot; also shown, is the undiagnosed fraction of the people that were infected and recovered without symptoms. This fraction, called the unaffected cases, is not measured or included in the data.}
   	\label{fig:Zambia}
  \end{center}
\end{figure}

\end{document}